\documentclass[aps,prb,showpacs,twocolumn,floats]{revtex4}
\usepackage{amssymb}

\usepackage{graphicx}
\usepackage{dcolumn}
\usepackage{bm}

\begin{document}

\bibliographystyle{prsty}
\author{Reem Jaafar, E. M. Chudnovsky, and D. A. Garanin}
\affiliation{Physics Department, Lehman College, City University
of New York \\ 250 Bedford Park Boulevard West, Bronx, New York
10468-1589, U.S.A.}
\date{\today}

\begin{abstract}
Local time-dependent theory of Einstein - de Haas effect is
developed. We begin with microscopic interactions and derive
dynamical equations that couple elastic deformations with internal
twists due to spins. The theory is applied to the description of
the motion of a magnetic cantilever caused by the oscillation of
the domain wall. Theoretical results are compared with a recent
experiment on Einstein - de Haas effect in a microcantilever.
\end{abstract}

\title{Dynamics of Einstein - de Haas Effect: Application to Magnetic Cantilever}

\maketitle

\section{Introduction}

Einstein - De Haas effect \cite{EdH} consists of the mechanical
rotation of a freely suspended body, caused by the change of its
magnetic moment. The latter can be induced by, e.g., the applied
magnetic field or by rapid warming. The Einstein - de Haas effect
is a direct consequence of the conservation of the total angular
momentum (spin + orbital). Consider, e.g., a solid made of $N$
atoms of magnetic moment ${\bf M} = \gamma_J {\bf J}$, where ${\bf
J} = {\bf S} + {\bf L}$ is the operator of the total angular
momentum of the atom (that includes spin ${\bf S}$ and orbital
moment ${\bf L}$), $\;\gamma_J = {g_J e}/({2mc})$ is the
gyromagnetic ratio for $J$, $e < 0$ is the charge of electron, and
$g_J = 1+ [2J(J+1)]^{-1}[J(J+1)+S(S+1) - L(L+1)]$ is the Lande
factor. Total angular momentum of the magnet suspended from a
string is a sum of $N\langle{\bf J}\rangle$ and the mechanical
orbital moment, $\mbox{\boldmath$\mathcal{L}$}$, due to the
rotation of the solid. If, for example, the solid, initially
non-magnetized and at rest, develops a macroscopic magnetic moment
$\mbox{\boldmath$\mathcal{M}$} = N\langle {\bf M} \rangle =
\gamma_J N \langle {\bf J} \rangle$, then the conservation law
requires that $N\langle {\bf J} \rangle +
\mbox{\boldmath$\mathcal{L}$} = 0$. This gives
$\mbox{\boldmath$\mathcal{L}$} =
-\mbox{\boldmath$\mathcal{M}$}/\gamma_J$, that is, the solid
begins to rotate on being magnetized.

Experiments on Einstein - de Haas effect and the related Barnett
effect \cite{Barnett} (generation of the magnetic moment by
mechanical rotation ), performed at the dawn of quantum physics,
provided first measurements of the gyromagnetic ratio for various
materials \cite{Frenkel}. Even today the Einstein - de Haas method
can still provide a more accurate value of $g_J$  as compared to
electron spin resonance and ferromagnetic resonance methods that
require precise knowledge of the effective magnetic field inside
the sample \cite{Bar-Scott}. Nevertheless fundamental questions
about the Einstein - de Haas effect remain unanswered. In
particular, the global conservation of the angular momentum does
not explain how the angular momentum is actually transferred from
individual atoms to the whole body. This question is clearly
related to the magnetic relaxation and decoherence at the atomic
level. The latter determine the width of para- and ferromagnetic
resonances, as well as functionality of spin-based qubits.
Advances in manufacturing and measuring of nanomechanical devices
promise to revive interest to the local dynamics of Einstein - de
Haas effect.

Our interest to this problem has been ignited by a recent
experiment performed at the NIST laboratory in Boulder, Colorado
\cite{NIST}. In that experiment a $50$nm permalloy film was
deposited onto a $200\mu$m$\times$$20\mu$m$\times$$0.6\mu$m
cantilever. The cantilever was placed inside a coil that generated
an ac magnetic field. Oscillation of the cantilever was measured
by a fiber optic interferometer positioned above the tip of the
cantilever. When the frequency of the ac field matched the
resonance frequency of the cantilever the amplitude of the
oscillations was about $3$nm. The data were analyzed within a
model that replaced the mechanical torque due to change in the
magnetization by the effect of the periodic force acting on the
fictitious point mass at the free end of the cantilever. Such an
approximation, while catching some features of the phenomenon, is
clearly insufficient for the study of the microscopic dynamics of
the Einstein - de Haas effect.

In this paper we will develop theoretical framework for the
description of the dynamics of the Einstein - de Haas effect, that
we will apply to the problem of magnetic cantilever. To make this
problem more transparent we shall assume (as is the case for many
magnetic solids) that the magnetism is of spin origin  and can be
described either by individual spins, ${\bf S}_i$, localized at
the atomic sites $i$, or by a continuous spin field ${\bf S}({\bf
r}, t)$. (Generalization to magnetism of spin and/or orbital
origin can be obtained through a straightforward re-definition of
the constants.) We shall derive general equations describing the
transfer of the spin angular momentum to the mechanical angular
momentum of the body. In the NIST experiment, the effect of the ac
magnetic field was likely the displacement of the domain wall
separating two magnetic domains inside the permalloy film. We
shall pay special attention to this case. The cantilever problem
will be solved by adding the internal torque due to the motion of
the domain wall to the equations of the elastic theory describing
the motion of the cantilever. The obtained dynamics of the
cantilever is rather rich and it allows a detailed comparison
between theory and experiment.

General theory of spin-rotation coupling will be studied in
Section II. Equations of elastic theory with internal twists due
to the dynamics of spins will be derived in Section III.
Mechanical motion of the magnetic cantilever will be studied in
Section IV. Suggestions for experiments will be given in Section
V.

\section{Microscopic theory of spin-rotation coupling}

\begin{figure}[ht]
\begin{center}
\includegraphics[width=50mm]{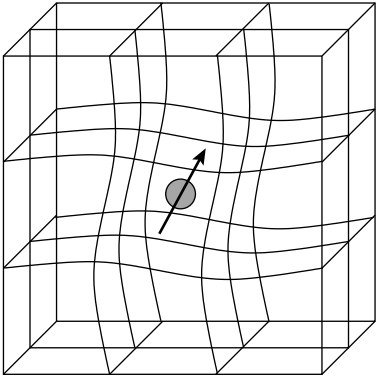}
\caption{Rotation of the crystal field due to local elastic twist
at the location of the spin.} \label{rotation}
\end{center}
\end{figure}

Spin-lattice interaction comes from magnetostriction and
spin-rotation coupling. Only the latter, however, is responsible
for the Einstein - de Haas effect. The most obvious effect of
local elastic twists comes from the dependence of the energy of a
spin on its orientation in the crystal - magnetic anisotropy. This
effect is due to spin-orbit interactions and is of relativistic
origin. It is described by the crystal-field Hamiltonian that can
be very generally written as
\begin{equation}\label{H1}
\hat{H}_{A} = \sum_jK_j^{\alpha\beta}S_j^{\alpha}S_j^{\beta} +
\sum_jL_j^{\alpha\beta\gamma\delta}S_j^{\alpha}S_j^{\beta}S_j^{\gamma}S_j^{\delta}
+ ...\,.
\end{equation}
Here the Greek letters denote Cartesian components of a
dimensionless spin vector ${\bf S}_j$ belonging to the site $j$ of
the crystal lattice. Tensors $K_j^{\alpha\beta}$,
$L_j^{\alpha\beta\gamma\delta}$, etc., describing magnetic
anisotropy, are defined in the coordinate frame
$\mathbf{e}_j^{(1,2,3)}$ that is rigidly coupled to the
locally-defined crystal axes, see Fig.\ \ref{rotation}.
Local rotation of the lattice is performed by the $(3\times 3)$
rotation matrix ${\mathbb{R}}_j$,
\begin{equation}
\mathbf{e}_j^{(1,2,3)}\rightarrow
{\mathbb{R}}_j\mathbf{e}_j^{(1,2,3)}\,. \label{rotation-e}
\end{equation}
It results in
\begin{eqnarray}\label{K,L}
& & K_j^{\alpha\beta}  \rightarrow {\mathbb{R}}_j^{\alpha \gamma
}{\mathbb{R}}_j^{\beta \delta }K_j^{\gamma\delta} \nonumber \\
& & L_j^{\alpha\beta\gamma\delta} \rightarrow
{\mathbb{R}}_j^{\alpha \eta }{\mathbb{R}}_j^{\beta \xi
}{\mathbb{R}}_j^{\gamma \theta
}{\mathbb{R}}_j^{\delta \epsilon } L_j^{\eta\xi\theta\epsilon}\\
& & ...\nonumber
\end{eqnarray}
For a small rotation at the site $j$ by an angle $\delta
\mathbf{\phi }_j$, one has
\begin{equation}
{\mathbb{R}}_j^{\alpha \beta }=\delta^{\alpha \beta }-\epsilon
^{\alpha \beta \gamma }\delta {\phi}_j^{\gamma}.
\label{rotation-matrix}
\end{equation}
We now notice that due to the rotational invariance of
$\hat{H}_{A}$, the
rotation of the local frame $(\mathbf{e}_j^{(1)},\mathbf{e}_j^{(2)},\mathbf{e}_j%
^{(3)})$ is equivalent to the rotation of the vector
$\mathbf{S}_j$ by the same angle in the opposite direction,
$\mathbf{S}\rightarrow {\mathbb{R}}^{-1}\mathbf{S}$. As is
known \cite{mes76}, this rotation can be equivalently performed by the $%
(2S+1)\times (2S+1)$ matrix in the spin space,
\begin{equation}
\mathbf{S}_j\rightarrow \hat{R}_j\mathbf{S}_j\hat{R}_j^{-1}\,,\qquad \hat{R}_j=e^{-i%
\mathbf{S}_j\cdot \delta \mathbf{\phi }_j}\,.
\end{equation}
Consequently, in the presence of rotations, $\hat{H}_{A}$ becomes
\cite{Fulde,CGS}
\begin{equation}
\hat{H}_{A}'=\hat{R}\hat{H}_{A}\hat{R}^{-1}\,, \label{Hfull}
\end{equation}
where
\begin{equation}\label{S-rotation}
\hat{R}=e^{-i\sum_j\mathbf{S}_j\cdot \delta \bm{\phi }_j}\,.
\end{equation}
In the linear order on $\delta \bm{\phi }_j$ one obtains
\begin{equation}\label{R-A}
\hat{R}\hat{H}_{A}\hat{R}^{-1}\mathbf{\cong
}\hat{H}_{A}\mathbf{+}\hat{H}_{R}, \qquad
\hat{H}_{R}=i\sum_j\left[ \hat{H}_{A},\mathbf{S}_j \right] \cdot
\delta \bm{\phi}_j \,. \label{Smallphi}
\end{equation}
By quantizing $\delta \bm{\phi}_j$ one can apply this Hamiltonian
to the study of rigid spin clusters and quantum dots
\cite{splaser,dot,Rabi}.

The total spin Hamiltonian $\hat{H}_S$ may include exchange
interaction, magnetostriction, Zeeman interaction and
dipole-dipole interaction. The dipole-dipole interaction is
usually the weakest one and will not be considered here. The
magnetostriction is local on spin. Consequently, it is transformed
by rotations the same way as the crystal field. The Zeeman
interaction of spins with the external magnetic field
$\mathbf{B}$,
\begin{equation}\label{Zeeman}
\hat{H}_{Z}=\sum_j\mathbf{b\cdot S}_j\,, \qquad {\bf b} \equiv
g\mu _{B}\mathbf{B} \label{HZDef}
\end{equation}
is not affected by rotations; $g$ being the gyromagnetic factor
for the spin. Here we take into account that the magnetic moment
due to spin, ${\cal{\bf M}} = -g\mu_B {\bf S}$, has direction
opposite to ${\bf S}$ because of the negative gyromagnetic ratio
for the electron. Finally, the exchange interaction,
\begin{equation}\label{ex}
\hat{H}_{\rm ex} = -\frac{1}{2}\sum_{ij}I_{ij}\, {\bf S}_i \cdot
{\bf S}_{j}\,,
\end{equation}
only depends on the local arrangement of spins that is not
affected by rotations. In the first order on $\delta \bm{\phi}_j$,
the generalization of Eq. (\ref{R-A}) is
\begin{equation}\label{S-P}
\hat{H}_{R} =  i\sum_j\left[ \hat{H}_{S},\mathbf{S}_j \right]
\cdot \delta \bm{\phi}_j - i\sum_j\left[ (\hat{H}_{Z}+\hat{H}_{\rm
ex}),\mathbf{S}_j\right]\cdot \delta \bm{\phi}_j \,.
\end{equation}
The last two terms appeared in Eq.\ (\ref{S-P}) because Zeeman
Hamiltonian (\ref{Zeeman}) and the exchange Hamiltonian
(\ref{ex}), that are included in $\hat{H}_{S}$, are independent
from local rotations. Consequently, one should subtract
$\hat{H}_{Z}$ and $\hat{H}_{\rm ex}$ from $\hat{H}_{S}$ when
computing the effect of rotations.

Let the total Hamiltonian of the system that incorporates all
couplings, including interactions with rotations, be $\hat{H}$. It
is clear that the difference between $[ \hat{H}_S,\mathbf{S}_j]$
and $[ \hat{H},\mathbf{S}_j]$ begins with the terms that are
linear on $\delta \bm{\phi}_j$. Thus, in the linear approximation
on $\delta \bm{\phi}_j$, we can replace $i[
\hat{H}_{S},\mathbf{S}_j]$ in Eq.\ (\ref{S-P}) with $i[
\hat{H},\mathbf{S}_j]=\hbar \dot{\bf S}_j$. Working out the
commutator with the Zeeman Hamiltonian in Eq.\ ({\ref{S-P}}) one
obtains
\begin{equation}\label{S-P-next}
\hat{H}_{R}= \sum_j\left(\hbar \dot{\bf S}_j + {\bf S}_j\times
{\bf b} - i\sum_j\left[ \hat{H}_{\rm ex},\mathbf{S}_j\right]
\right) \cdot \delta {\bm {\phi}}_j\,.
\end{equation}

\section{Elastic theory with internal twists due to spin-rotation coupling}

Our approach to Einstein - de Haas effect is based upon Eq.\
(\ref{S-P-next}). To apply this equation to long-wave torsional
deformations of the body we shall write $\delta \bm{\phi}_j$ in
terms of the displacement field of the elastic theory ${\bf
u}({\bf r},t)$ \cite{LL},
\begin{equation}\label{phi-u}
\delta \mathbf{\phi (\mathbf{r})=}\frac{1}{2}\nabla \times \mathbf{u}(%
\mathbf{r})  \,,
\end{equation}
and replace ${\bf S}_j$ by the spin density ${\bf S}({\bf r}, t)$.
The classical energy of the body then becomes
\begin{equation}\label{H}
{\cal{H}} = {\cal{H}}_{S} + {\cal{H}}_{E} + {\cal{H}}_{R}\,.
\end{equation}
Here ${\cal{H}}_{E}$ is the elastic energy of the body written in
terms of ${\bf u}({\bf r}, t)$ while ${\cal{H}}_S = \langle H_S
\rangle$ includes exchange, anisotropy, Zeeman and dipolar
energies, magnetostriction, etc., written in terms of ${\bf
S}({\bf r}, t)$ and ${\bf u}({\bf r}, t)$ \cite{book}. The last
term in Eq.\ (\ref{H}) follows from equations (\ref{S-P-next}) and
(\ref{phi-u}),
\begin{equation}\label{int}
{\cal{H}}_{R} = \frac{1}{2}\int d^3r \,\left[\hbar \dot{\bf S} +
{\bf S}\times ({\bf b} + {\bf b}_{\rm ex})\right] \cdot (\nabla
\times \mathbf{u})\,,
\end{equation}
where \cite{book}
\begin{equation}\label{b-ex}
{\bf b}_{\rm ex} = \frac{\delta {\cal{H}}_{\rm ex}}{\delta {\bf
S}} = -I_{\alpha\beta}\frac{{\partial}^2 {\bf
S}}{{\partial}r_{\alpha}{\partial}r_{\beta}}
\end{equation}
and
\begin{equation}
I_{\alpha\beta} =  \frac{1}{2} \sum_{j}
I_{ij}(r_i^{\alpha}-r_j^{\alpha})(r_i^{\beta}-r_j^{\beta})\,.
\end{equation}

The dynamical equation for the displacement field is \cite{LL}
\begin{equation}\label{elastic}
\rho \frac{\partial^2 u_{\alpha}}{\partial t^2} = \frac{\partial
\sigma_{\alpha \beta}}{\partial x_{\beta}} \,,
\end{equation}
where $\sigma_{\alpha \beta} = {\delta {\cal{H}}}/\delta e_{\alpha
\beta}$ is the stress tensor, $e_{\alpha \beta} = \partial
u_{\alpha}/\partial x_{\beta}$ is the strain tensor, and $\rho$ is
the mass density of the material. The stress tensor can be divided
into two parts, $\sigma_{\alpha \beta} = \sigma^{(\rm{M})}_{\alpha
\beta} + \sigma^{(\rm{R})}_{\alpha \beta}$, with
\begin{equation}
\sigma^{(\rm{M})}_{\alpha \beta} = \frac{\delta ({\cal{H}}_{S} +
{\cal{H}}_{E} )}{\delta e_{\alpha \beta}}
\end{equation}
and
\begin{equation}
\sigma^{(\rm{R})}_{\alpha \beta} = \frac{\delta
{\cal{H}}_{R}}{\delta e_{\alpha \beta}}\,.
\end{equation}
Here $\sigma^{(\rm{M})}_{\alpha \beta}$ is the mechanical part of
the stress tensor, e.g., the part coming from the elastic
properties of the cantilever plus magnetostriction, while
$\sigma^{(\rm{R})}_{\alpha \beta}$ is the part of the stress
tensor produced by the internal rotations due to spins,
\begin{equation}\label{sigma}
\sigma^{(\rm{R})}_{\alpha \beta} = -\frac{1}{2}\epsilon_{\alpha
\beta \gamma}\left\{\hbar \dot{S}_{\gamma} + [{\bf S}\times ({\bf
b} + {\bf b}_{\rm ex})]_{\gamma}\right\}\,.
\end{equation}
Notice that, contrary to the symmetric stress tensor
($\sigma_{\alpha \beta} = \sigma_{\beta \alpha}$) used by the
conventional elastic theory, $\sigma^{(\rm{R})}_{\alpha \beta}$ is
antisymmetric. The immediate consequence of that is a non-zero
torque,
\begin{equation}\label{K}
dK^{(\rm{R})}_{\alpha \beta} = (\sigma_{\alpha \beta} -
\sigma_{\beta \alpha})d^3r\,,
\end{equation}
acting on the volume element $d^3r$. Such torques, neglected by
the conventional theory of elasticity, are responsible for the
Einsten - de Haas effect.

Equations (\ref{elastic}) - (\ref{sigma}) allow one to obtain the
general dynamical equation of the elastic theory that accounts for
local internal forces due to the dynamics of spins in a
ferromagnet:
\begin{equation}\label{torque}
\rho \frac{\partial^2 u_{\alpha}}{\partial t^2} - \frac{\partial
\sigma^{(\rm{M})}_{\alpha \beta}}{\partial x_{\beta}} = f^{({\rm
R})}_{\alpha}\,,
\end{equation}
where
\begin{equation}\label{force}
{\bf f}^{({\rm R})} = - \frac{1}{2}{\bm \nabla}\times \left[\hbar
\dot{\bf S} + {\bf S}\times ({\bf b} + {\bf b}_{\rm ex})\right]\,.
\end{equation}
Let us check that these equations conserve the total angular
momentum (spin + orbital). Writing the total angular momentum due
to the spins and the crystal as
\begin{equation}\label{J}
{\bf J} = \int d^3r \left[ \hbar{\bf S} + \rho\left({\bf r} \times
\dot{\bf u}\right)\right]\,,
\end{equation}
one obtains the following equation for the time derivative of the
$\alpha$-th component of ${\bf J}$,
\begin{eqnarray}\label{J-dot}
\dot{J}_{\alpha} & = &\int d^3 r\left[ \hbar\dot{S}_{\alpha} +
\epsilon_{\alpha\beta\gamma}r_{\beta}\left(\rho\ddot{u}_{\gamma}\right)\right]\nonumber
\\
 & = &\int d^3 r\left\{ \hbar\dot{S}_{\alpha} +
\epsilon_{\alpha\beta\gamma}r_{\beta}\nabla_{\delta}\sigma^{(\rm{M})}_{\gamma
\delta}\right.\nonumber \\
&-& \left.\frac{1}{2}\left[r_{\beta}\nabla_{\alpha}
 -\left({\bf r}\cdot{\bm
\nabla}\right)\delta_{\alpha\beta}\right]\times \right.\nonumber \\
& & \left.\left[\hbar \dot{S}_{\beta} + ({\bf S}\times {\bf b}
)_{\beta}
-\epsilon_{\beta\gamma\delta}I_{\epsilon\eta}S_{\gamma}\nabla_{\epsilon}\nabla_{\eta}S_{\delta}
\right]\right\},
\end{eqnarray}
where we have used equations (\ref{torque}), (\ref{force}) and
(\ref{b-ex}). If one prohibits transfer of spin angular momentum
through the surface, integration by parts with account of the
symmetry of $\sigma^{(\rm{M})}_{\alpha \beta}$ and
$I_{\alpha\beta}$ gives
\begin{equation}\label{J-eq}
\dot{\bf J} = {\bf K}^{(M)}+ {\bf K}^{(R)}\,,
\end{equation}
with
\begin{equation}
{K}_{\alpha}^{(M)} = \int d
A_{\delta}\left[\epsilon_{\alpha\beta\gamma}r_{\beta}\sigma^{(M)}_{\gamma\delta}\right]\label{KM}
\end{equation}
and
\begin{equation}
{\bf K}^{(R)} = \int d^3r \left({\bf b}\times{\bf
S}\right)\label{KR}\,.
\end{equation}
Here ${\bf K}^{(M)}$ is the external mechanical torque applied to
the surface of the body ${\bf A}$, while ${\bf K}^{(R)}$ is the
volume spin torque due to the external magnetic field. Thus, in
accordance with our expectation, when external forces are absent,
Eq.\ (\ref{torque}) conserves the total angular momentum,
$\dot{\bf J}=0$.

If the spin-lattice interaction was absent there would be no
deformation induced by the dynamics of spins. This condition
provides another check of the validity of equations (\ref{torque})
and (\ref{force}). In the absence of dissipation the spin field
satisfies \cite{book}
\begin{equation}\label{LL-eq}
\hbar \dot{\bf S} = -{\bf S} \times {\bf b}_{\rm eff}\,
\end{equation}
where ${\bf b}_{\rm eff} = {\delta {\cal{H}}}/{\delta {\bf S}}$ is
the effective field that can be presented as ${\bf b}_{\rm eff} =
{\bf b} + {\bf b}_{\rm ex} + {\bf b}'$. Here ${\bf b}'$ is
determined by the the spin-lattice coupling. Its main part is
usually the anisotropy field, ${\bf b}_A = {\delta
{\cal{H}}_A}/{\delta {\bf S}}$. Equations (\ref{torque}) and
(\ref{LL-eq}) then show that Zeeman and exchange interactions
alone do not provide any force on the body. This is in accordance
with the fact that spins should couple to the lattice in order to
produce such a force. Dissipation can be incorporated into the
problem by adding standard damping terms to the elastic equation
(\ref{torque}) and Landau-Lifshitz equation (\ref{LL-eq}).

\section{Dynamics of magnetic cantilever}

Among many problems involving internal forces due to spins, Eq.\
(\ref{torque}) can be used for computation of the elastic motion
of a magnetic cantilever. For example, in the case of the motion
of a domain wall inside the cantilever, one substitutes the known
domain wall solution for ${\bf S}({\bf r}, t)$ into the right-hand
side of Eq.\ (\ref{torque}), while the left-hand side follows from
the elastic theory of cantilever in the absence of spins
\cite{LL}. Sudden increase of the external magnetic field should
result in the domain wall sweeping the cantilever, thus providing
a source of deformation during a finite time. Application of a
harmonic ac magnetic field, as in the NIST experiment, should lead
to the oscillation of the position of the domain wall inside the
cantilever.

\begin{figure}[ht]
\begin{center}
\vspace{-0.8cm}
\includegraphics[width=70mm,angle=-90]{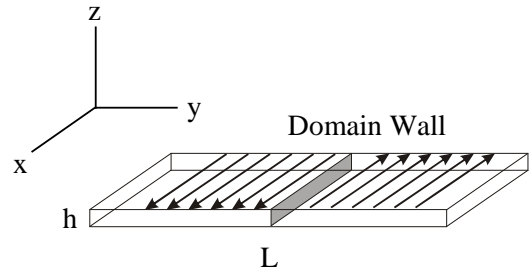}
\vspace{-3cm} \caption{Geometry considered in the paper.}
\label{NIST}
\end{center}
\end{figure}

The geometry of the problem is shown in Fig. \ref{NIST}.
The cantilever of length $L$, parallel to the $Y$-direction, is
magnetized in the $X$-direction. The $y = 0$ end of the cantilever
is attached to the holder while the $y = L$ end is free. We are
interested in small displacements of the cantilever in the
$Z$-direction, $u_z(y,t)$, caused by the time-dependent external
magnetic field. Vectors ${\bf B}$ and ${\bf S}$ are assumed to lie
in the $XY$ plane. The latter property of the magnetization is
common for thin films. It is easy to see that in this case the
terms proportional to ${\bf S}\times{\bf b}$ and ${\bf S}
\times{\bf b}_{ex}$ in the right-hand side of Eq. (\ref{torque})
give zero contribution to the $Z$-component of the elastic
equation. Adding the term proportional to $\dot{\bf S}$ to the
right-hand side of the conventional equation of motion for a
cantilever \cite{LL}, one obtains from Eq.\ (\ref{torque})
\begin{equation}\label{elastic-S}
\rho \frac{\partial^2 u_z}{\partial t^2} +
\frac{h^2E}{12(1-\sigma^2)}\frac{\partial^4 u_z}{\partial y^4} =
\frac{\hbar}{2}\frac{\partial}{\partial y}\frac{\partial
}{\partial t}S_{x}(y,t)\,,
\end{equation}
where $h$ is the thickness of the cantilever in the $Z$-direction,
$E$ is the Young's modulus, and $\sigma$ is the Poisson
coefficient, $-1 < \sigma < 1/2$.

If the magnetization of the cantilever was rotating uniformly in
space, then according to Eq. (\ref{elastic-S}) the force from
spins would only act on the free end of the cantilever where the
magnetization has a discontinuity. However, for a soft magnetic
material like permalloy, deposited on a cantilever that is large
compared to the dimensions of a monodomain particle, the change in
the magnetization should occur through the motion of a domain
wall. For that reason we shall describe the magnetic state of the
cantilever by two domains separated by the domain wall at $y =
y_0(t)$. Then $S_{x}(y,t)$ is given by the domain wall solution
centered at $y = y_0(t)$,
\begin{equation}
S_{x}(y,t) = S_{\rm dw}\left[y - y_0(t)\right]\,.
\end{equation}
In the absence of the dc magnetic field this gives
\begin{eqnarray}\label{Sdw}
& & \rho \frac{\partial^2 u_z}{\partial t^2} +
\frac{h^2E}{12(1-\sigma^2)}\frac{\partial^4 u_z}{\partial y^4} =
\nonumber \\
& & - \frac{\hbar}{2}\left(\frac{d
y_0}{dt}\right)\frac{\partial^2}{\partial y^2}S_{\rm dw}\left[y -
y_0(t)\right]\,.
\end{eqnarray}
For the dissipative motion of the domain wall the speed of the
wall is proportional to the field. When the ac magnetic field, $B
= B_0\cos(\omega t)$, is applied in the $X$-direction, one has
$y_0(t) = y_0(0) + a\sin(\omega t)$, where $a < L$ is the
amplitude of the oscillations around $y_0(0)$. The domain wall is
given by $S_{\rm dw}(y,t) = S_0 F\left[y - y_0(t)\right]$ where
$S_0$ is a constant spin density and $F$ changes from $-1$ to $+1$
as one crosses the wall. Note the connection of $S_0$ to the
magnetization, $M_0 = g\mu_B S_0$.

It is convenient to switch to dimensionless variables,
\begin{equation}\label{bar}
\bar{u}_z = \frac{u_z}{L}\,, \quad \bar{y} = \frac{y}{L}\,, \quad
\bar{t} = t\nu\,, \quad \nu \equiv
\sqrt{\frac{Eh^2}{12\rho(1-\sigma^2)L^4}}\,,
\end{equation}
where $L$ is the length of the cantilever and $\nu$ determines the
scale of the eigenfrequencies of its vertical oscillations,
$u_z(y,t)$. In terms of these variables Eq.\ (\ref{Sdw}) becomes
\begin{equation}\label{eq-bar}
\frac{\partial^2 \bar{u}_z}{\partial \bar{t}^2} + \frac{\partial^4
\bar{u}_z}{\partial \bar{y}^4} = -\epsilon \left(\frac{d
\bar{y}_0}{d\bar{t}}\right)\frac{\partial^2 F}{\partial
\bar{y}^2}\,,
\end{equation}
where
\begin{equation}\label{epsilon}
\epsilon = \frac{\hbar S_0}{2\rho L^2\nu}= \frac{\hbar
S_0}{2\rho}\sqrt{\frac{12\rho(1-\sigma^2)}{Eh^2}}
\end{equation}
is a dimensionless parameter that does not depend on the length of
the cantilever $L$.  By order of magnitude, $\epsilon \sim
\hbar/(Msh)$, where $M \sim \rho/S_0$ is the mass of the material
per spin 1/2 and $s \sim \sqrt{E/\rho}$ is the speed of sound. It
is easy to see that $\epsilon$ is a small parameter that can
hardly exceed $0.01$ even for the smallest cantilevers.

For the given function $\bar{y}_0(\bar{t})$ that describes the
motion of the domain wall, Eq.\ (\ref{eq-bar}) has to be solved
with the following boundary conditions:
\begin{eqnarray}\label{boundary}
& & \bar{u}_z = 0\,,\; \frac{\partial \bar{u}_z}{\partial \bar{y}}
= 0\;\;\; {\rm at}\;\;\; \bar{y} = 0\,, \nonumber \\
& &
\frac{\partial^2 \bar{u}_z}{\partial \bar{y}^2} = 0\,,\;
\frac{\partial^3 \bar{u}_z}{\partial \bar{y}^3} = 0\;\;\; {\rm
at}\;\;\; \bar{y} = 1\,.
\end{eqnarray}
The first two conditions correspond to the absence of displacement
and the absence of bending of the cantilever at the fixed end,
while the last two conditions correspond to the absence of torque
and force, respectively, at the free end \cite{LL}.

\begin{figure}[ht]
\begin{center}
\includegraphics[width=60mm, angle = -90]{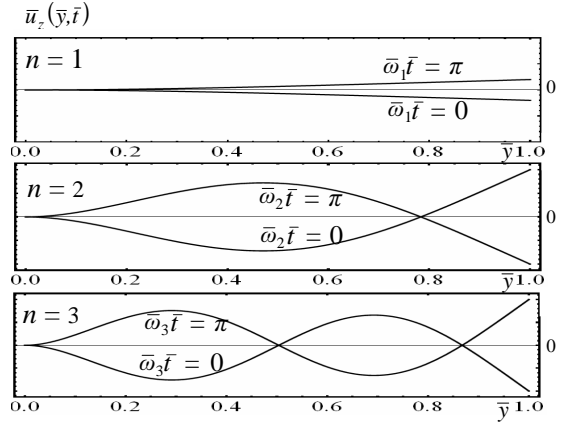}
\caption{Profiles of the oscillating cantilever at different
moments of time for $n = 1,2,3$. } \label{Free}
\end{center}
\end{figure}

For the free oscillations of the cantilever, $\epsilon = 0$, one
writes
\begin{equation}\label{free}
\bar{u}_z(\bar{y}, \bar{t}) =
\bar{u}(\bar{y})\cos(\bar{\omega}\bar{t})\,.
\end{equation}
Substitution into Eq.\ (\ref{eq-bar}) with $\epsilon = 0$ then
gives
\begin{equation}\label{eq-u}
\frac{\partial^4 \bar{u}}{\partial \bar{y}^4} - \kappa^4\bar{u} =
0\,, \qquad \kappa^2 \equiv \bar{\omega}\,.
\end{equation}
The general solution of this equation is
\begin{equation}\label{general}
\bar{u}(\bar{y}) = A\cos(\kappa \bar{y}) + B\sin(\kappa \bar{y}) +
C \cosh(\kappa \bar{y}) + D \sinh(\kappa \bar{y})\,,
\end{equation}
where $A,B,C,D$ are constants of integration. With the help of the
first, second, and fourth boundary conditions (\ref{boundary}) one
obtains
\begin{equation}
C=-A\,, \quad D = -B\,, \quad B = \frac{\sin\kappa -
\sinh\kappa}{\cos\kappa + \cosh\kappa}A\,.
\end{equation}
Substitution of these expressions into Eq.\ (\ref{general}) gives
up to a normalization factor
\begin{eqnarray}\label{eigenfunction}
& & \bar{u}(\bar{y}) = (\cos\kappa + \cosh\kappa)\left[\cos(\kappa
\bar{y}) - \cosh(\kappa
\bar{y})\right] \nonumber \\
& & + (\sin\kappa - \sinh\kappa)\left[\sin(\kappa \bar{y})-
\sinh(\kappa \bar{y})\right]\,.
\end{eqnarray}
The third of the boundary conditions (\ref{boundary}) provides
equation,
\begin{equation}\label{modes}
\cos\kappa\cosh\kappa + 1 = 0\,,
\end{equation}
for the frequencies of the normal modes of the cantilever,
$\bar{\omega}_n = \kappa^2_n$ (measured in the units of $\nu$ of
Eq.\ (\ref{bar})). Fundamental (minimal) frequency is
$\bar{\omega}_1 \approx 3.516$. The next two frequencies are
$\bar{\omega}_2 \approx 22.03$ and $\bar{\omega}_3 \approx 61.70$.
The profiles of the oscillations of the cantilever for three
normal modes ($n=1,2,3$) are shown in Fig. \ref{Free}.

We shall now turn to the forced oscillations of the cantilever due
to motion of the domain wall. We first neglect dissipation and
write for the displacement
\begin{equation}\label{forced}
\bar{u}_z(\bar{y}, \bar{t}) =
\sum_mR_m(\bar{t})\bar{u}_m(\bar{y})\,,
\end{equation}
where $R_m(t)$ are functions of time to be determined and $
\bar{u}_m(\bar{y})$ are normalized eigenfunctions
(\ref{eigenfunction}) of the free cantilever that correspond to
eigenvalues $\kappa_m$ given by Eq.\ (\ref{modes}),
\begin{equation}\label{norm}
\int_0^1dy\,\bar{u}_m(\bar{y})\bar{u}_n(\bar{y}) = \delta_{mn}\,.
\end{equation}
Substitution of Eq.\ (\ref{forced}) into Eq.\ (\ref{eq-bar}) gives
\begin{equation}
\sum_m\left(\frac{d^2R_m}{d\bar{t}^2} + \bar{\omega}_m^2
R_m\right)\bar{u}_m(\bar{y}) = -\epsilon \left(\frac{d
\bar{y}_0}{d\bar{t}}\right)\frac{\partial^2 F}{\partial
\bar{y}^2}\,,
\end{equation}
where we have used Eq.\ (\ref{eq-u}). Multiplying both parts of
this equation by $\bar{u}_n(\bar{y})$ and integrating over
$\bar{y}$ from $0$ to $1$ with account of Eq.\ (\ref{norm}), one
obtains linear second-order differential equation for
$R_n(\bar{t})$,
\begin{equation}\label{R-eq}
\frac{d^2R_n}{d\bar{t}^2} + \bar{\omega}_n^2 R_n = -\epsilon
\left(\frac{d
\bar{y}_0}{d\bar{t}}\right)\int^1_{0}d\bar{y}\frac{\partial^2
F}{\partial \bar{y}^2}\bar{u}_n(\bar{y})\,.
\end{equation}
When the width of the domain wall is small compared to the length
of the cantilever, the first derivative of $F$ can be replaced by
the $\delta$-function,
\begin{equation}\label{delta}
\frac{\partial F}{\partial \bar{y}} = 2\delta[\bar{y} -
\bar{y}_0(\bar{t})]\,.
\end{equation}
In this case, integrating by parts in the right-hand side of Eq.\
(\ref{R-eq}), one obtains
\begin{eqnarray}
 \frac{d^2R_n}{d\bar{t}^2} + \bar{\omega}_n^2 R_n & = & 2\epsilon
\left(\frac{d \bar{y}_0}{d\bar{t}}\right)\left(\frac{d
\bar{u}_n}{d\bar{y}}\right)_{\bar{y}= \bar{y}_0(\bar{t})}
\nonumber \\
& = &-2\epsilon\frac{d}{d\bar{t}}\,
\bar{u}_n\left[\bar{y}_0(\bar{t})\right]\,.
\end{eqnarray}
Dissipation can be included into the problem by adding the first
time derivative of $R_n$ to this equation. This results in a
conventional problem of damped oscillations induced by a periodic
force:
\begin{equation}\label{R-dis}
\frac{d^2R_n}{d\bar{t}^2}+
\frac{\bar{\omega}_n}{Q_n}\frac{dR_n}{d\bar{t}} + \bar{\omega}_n^2
R_n = -2\epsilon\frac{d}{d\bar{t}}\,
\bar{u}_n\left[\bar{y}_0(\bar{t})\right]\,.
\end{equation}
Here $Q_n$ is the quality factor of the oscillations of the
cantilever at the eigenfrequency $\bar{\omega}_n$.

The most interesting case is when the position of the domain wall,
\begin{equation}\label{DW-oscillation}
\bar{y}_0(\bar{t}) = \bar{b} + \bar{a}\sin(\bar{\omega}\bar{t})\,,
\end{equation}
oscillates at a frequency $\bar{\omega}$ that is close to one of
the resonant frequencies of the cantilever $\bar{\omega}_n =
\kappa^2_n$. To solve Eq.\ (\ref{R-dis}) we write $u_n$ and $R_n$
as Fourier series,
\begin{equation}\label{Fourier}
u_n(\bar{t}) = \sum_{k =
-\infty}^{\infty}u_k^{(n)}e^{ik\bar{\omega}\bar{t}}\,, \qquad
R_n(\bar{t}) = \sum_{k =
-\infty}^{\infty}r_k^{(n)}e^{ik\bar{\omega}\bar{t}}\,.
\end{equation}
Substitution into Eq.\ (\ref{R-dis}) gives
\begin{equation}
r_k^{(n)} = \frac{-2i\epsilon
k\bar{\omega}u_k^{(n)}}{\bar{\omega}_n^2-k^2\bar{\omega}^2
+\frac{ik\bar{\omega}\bar{\omega}_n}{Q_n}}\,,
\end{equation}
where
\begin{equation}
u_k^{(n)}(\bar{a},\bar{b}) = \frac{1}{2\pi}\int_0^{2\pi}d \xi
e^{-ik\xi} \bar{u}_n\left(\bar{b}+\bar{a}\sin\xi\right)\,.
\end{equation}
Writing $u_k^{(n)}$ as $u_k^{(n)} =
|u_k^{(n)}|\exp{[i\gamma_k^{(n)}]}$ one obtains the following
expressions for the amplitude $|r_k^{(n)}|$ and phase
$\delta_k^{(n)}$ of the $k$-th harmonic of $R_n(t)$:
\begin{equation}\label{amplitude}
|r_k^{(n)}| = \frac{2\epsilon
k\bar{\omega}|u_k^{(n)}|}{\sqrt{(k^2\bar{\omega}^2 -
\bar{\omega}_n^2)^2
+\left(\frac{k\bar{\omega}\bar{\omega}_n}{Q_n}\right)^2}}
\end{equation}
\begin{equation}\label{phase}
\delta_k^{(n)} = \gamma_k^{(n)} -\frac{\pi}{2}+
\arctan\left[\frac{k\bar{\omega}\bar{\omega}_n}{Q_n(k^2\bar{\omega}^2
-\bar{\omega}_n^2)}\right]\,.
\end{equation}

\begin{figure}[ht]
\begin{center}
\includegraphics[width=60mm,angle=-90]{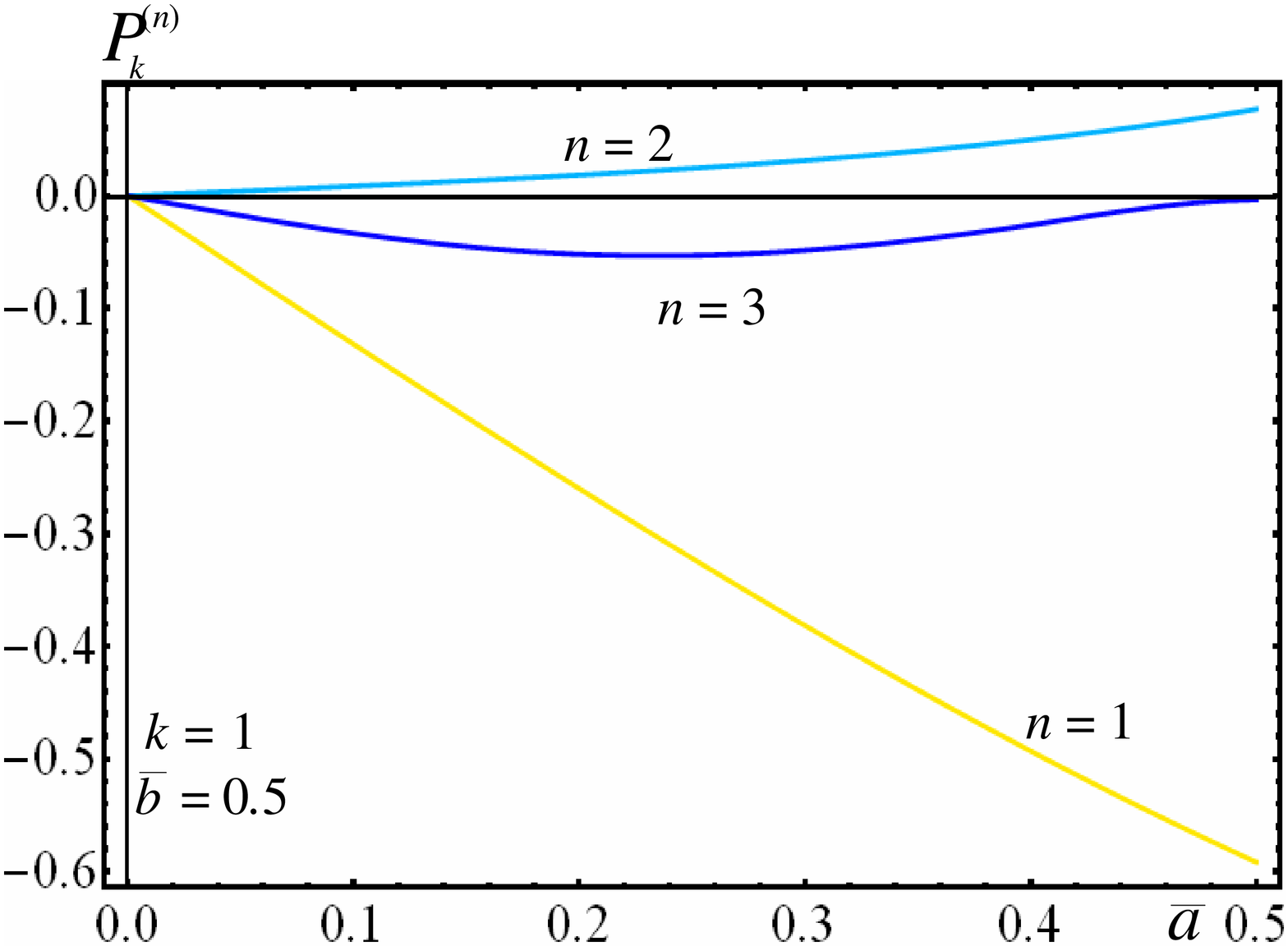}
\includegraphics[width=60mm,angle=-90]{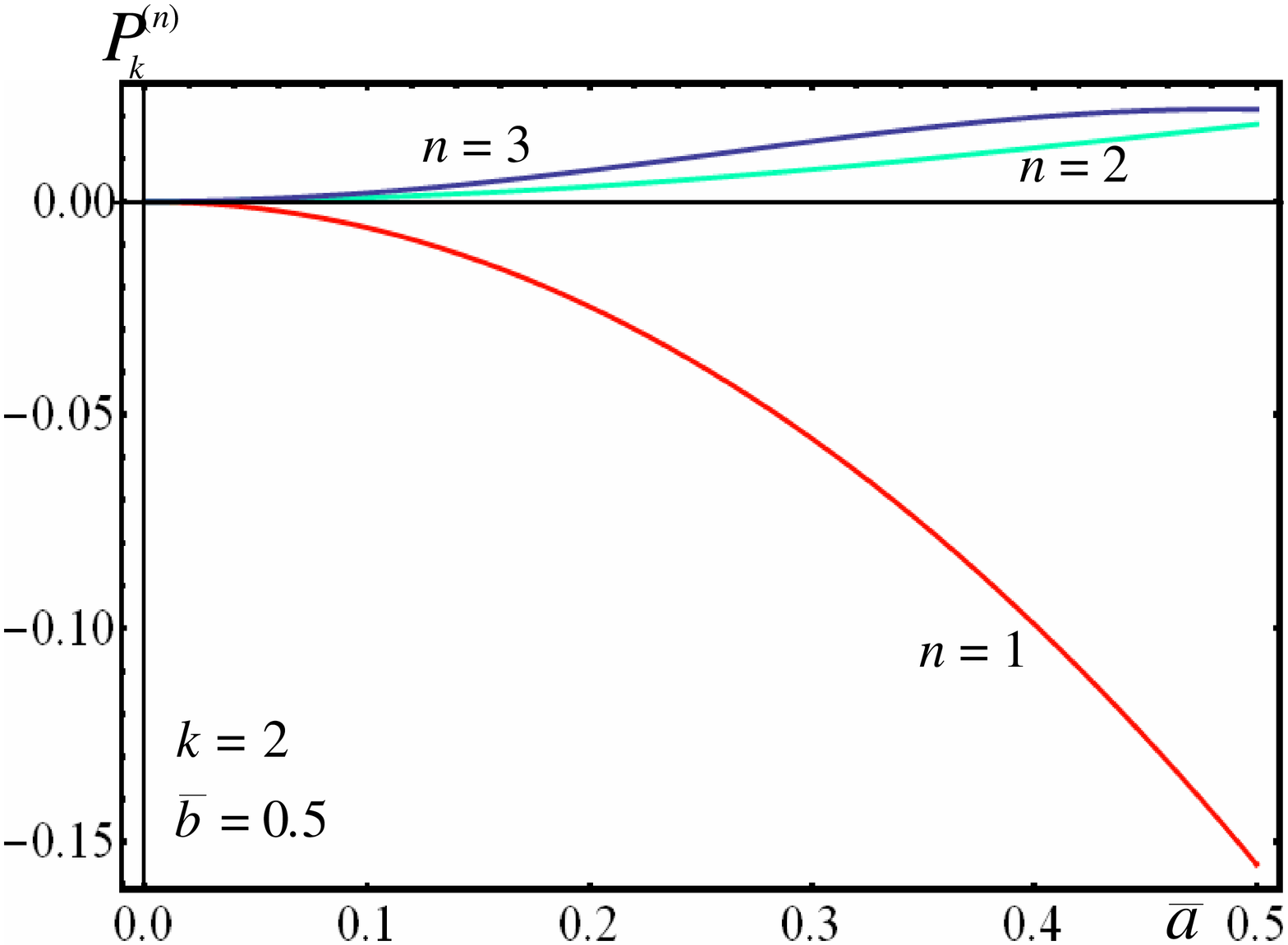}
\caption{Color on line: The dependence of
$P_k^{(n)}(\bar{a},\bar{b})$ on the amplitude of the oscillations
of the domain wall whose equilibrium position is in the middle of
the cantilever: a) $k =1$, b) $k =2$.} \label{P-a}
\end{center}
\end{figure}

\begin{figure}[ht]
\begin{center}
\includegraphics[width=60mm,angle=-90]{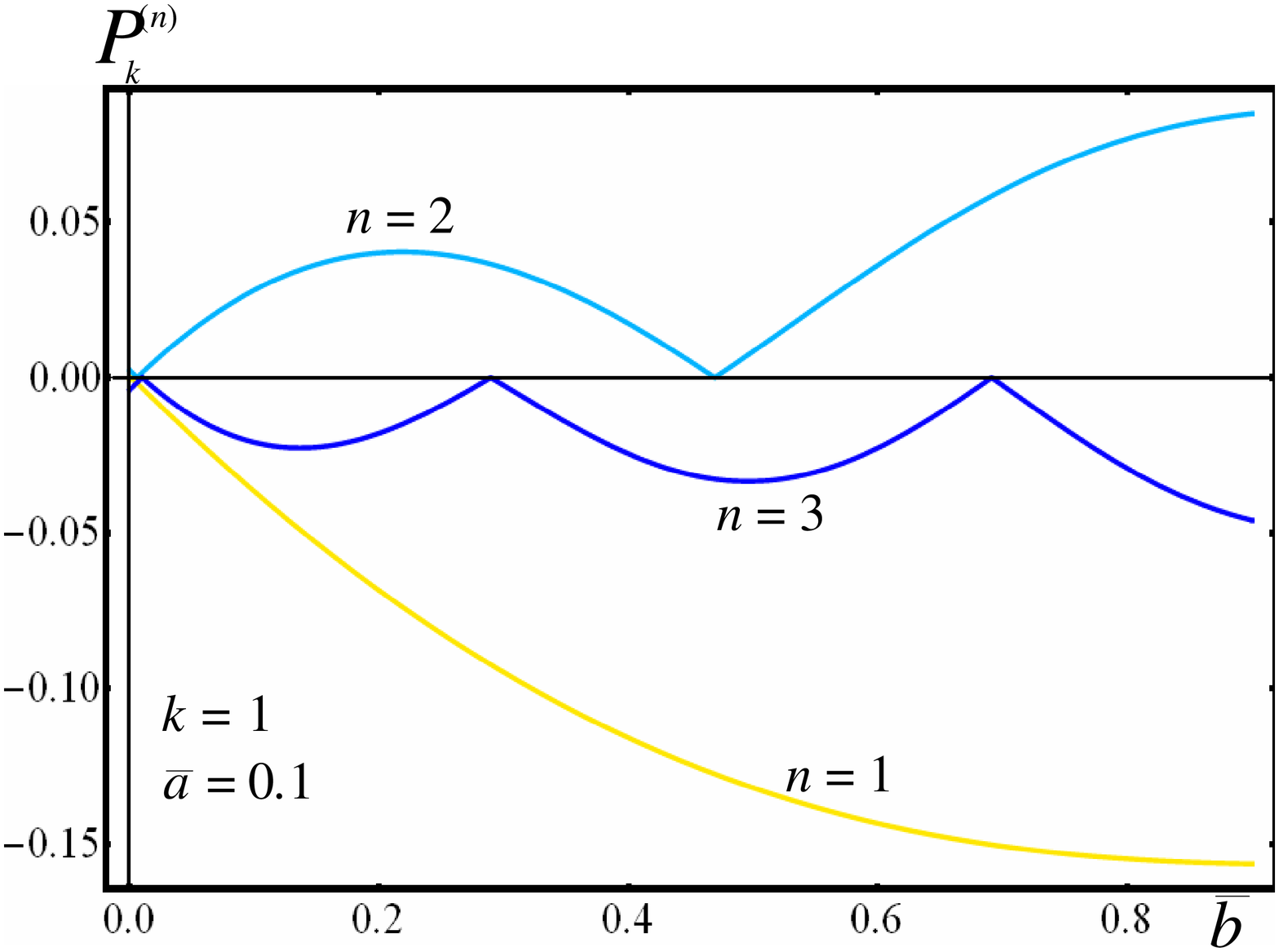}
\includegraphics[width=60mm,angle=-90]{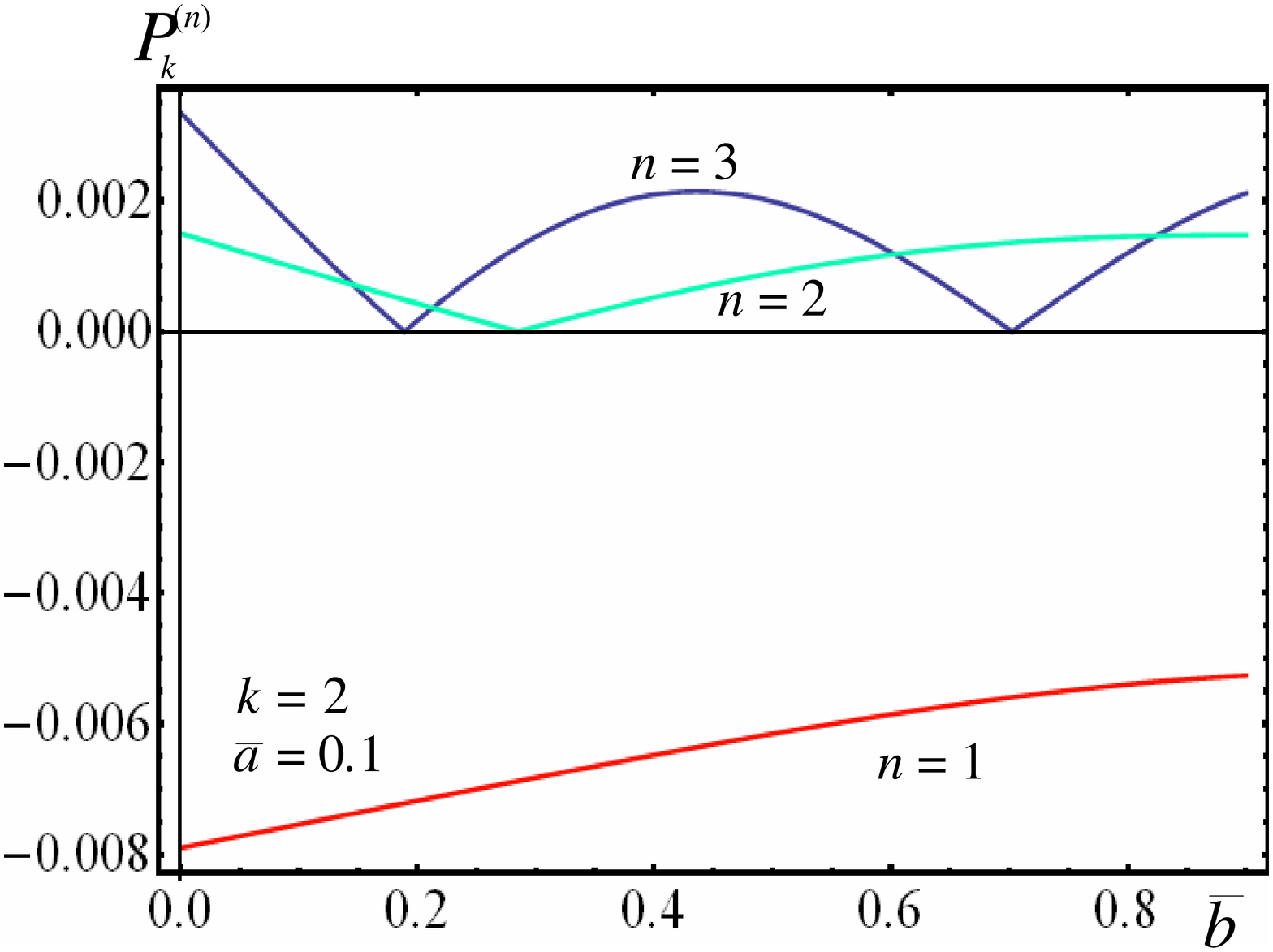}
\caption{Color on line: The dependence of
$P_k^{(n)}(\bar{a},\bar{b})$ on the equilibrium position of the
domain wall for $n=1,2,3$ and $k =1,2$.} \label{P-b}
\end{center}
\end{figure}

According to Eq.\ (\ref{amplitude}), resonances occur at
frequencies $\omega = \omega_n/k$ that are independent of damping.
At $\omega = \omega_n/k$ the maximum displacement of the free end
of the cantilever is given by
\begin{equation}\label{max}
\bar{u}_z(1) = \epsilon Q_n P_k^{(n)}(\bar{a},\bar{b})\,,
\end{equation}
where
\begin{equation}
P_k^{(n)}(\bar{a},\bar{b}) =
\frac{2\bar{u}_n(1)}{\bar{\omega}_n}\left|u_k^{(n)}(\bar{a},\bar{b})\right|\,.
\end{equation}
The dependence of $P_k^{(n)}$ on $\bar{a}$ and $\bar{b}$ for
various $k$ and $n$ is illustrated in Fig. \ref{P-a}  and Fig.
\ref{P-b}.
Note the non-monotonic dependence of the amplitude of the free end
on the equilibrium position of the domain wall for $n = 2$ and $n
= 3$. It is due to the profile of the normal modes of the
cantilever shown in Fig. \ref{Free}. When equilibrium position of
the wall coincides with the antinode, the effect of the
oscillation of the wall on the cantilever is minimal.

\section{Suggestions for experiment}

Expressions derived in this paper provide the framework for
theoretical analysis of the experimental data on Einstein - de
Haas effect. To illustrate applications of the theory we have
derived rigorous formulas for the mechanical motion of a magnetic
cantilever, induced by the motion of a domain wall when the
cantilever is placed in the ac magnetic field. In our theory we
assumed that the entire volume of the cantilever was magnetic. The
formulas can be easily adjusted, however, to the situation when
the magnetic layer has thickness $ph < h$, as was the case in the
NIST experiment. In this case the strength of the source in the
right hand side of Eq.\ (\ref{eq-bar}) reduces by the factor $p$.
Consequently, one should replace $\epsilon$ in the above formulas
with $\epsilon_p = p \epsilon < \epsilon$.

Accurate comparison between theory and experiment requires precise
knowledge of the mechanism by which the magnetic moment is
changing. If it is due to the motion of the domain wall, as we
believe was the case in the NIST experiment \cite{NIST}, then one
needs to know the initial equilibrium position $b$ and the
amplitude of the oscillations of the wall $a$. The parameter $b$
can be controlled by a weak dc magnetic field, while $a$ can be
controlled by the amplitude of the ac field. It is also desirable
to excite various harmonics $\omega_n/k$ and to identify the
fundamental frequency $\omega_1$. This would allow one to obtain
the value of the parameter $\nu$ in equations (\ref{bar}) and
(\ref{epsilon}). If the magnetization and the $g$-factor are
known, the ratio $S_0/\rho$ in the first of Eq.\ (\ref{epsilon})
can be computed with good accuracy. The precision with which the
parameter $\epsilon$ can be determined will then depend on the
knowledge of the length of the cantilever $L$. Alternatively,
$\epsilon$ can be extracted from experiment and used to obtain the
spin density $S_0$. If the magnetization $M_0$ is known this would
allow one to obtain the gyromagnetic factor $g = M_0/(\mu_BS_0)$.

In the NIST experiment the fraction of the magnetic material $p$
was close to $1/12$ while the dimensions of the cantilever were $L
= 2 \times 10^{-4}$m, $h = 6 \times 10^{-7}$m. This gives
$\epsilon_p \sim 10^{-7}$. If $a$ is comparable to $L$, then
according to Eq.\ (\ref{max}), the deflection of the free end of
the cantilever at the fundamental frequency $\omega = \omega_1$
must be of order $\epsilon_p Q_1 L$. The observed deflection in
the nanometer range then corresponds to $Q_1 \sim 100$. We should
notice in this connection that the effect could be stronger for a
cantilever with a higher quality factor. As a matter of fact the
quality factors as high as $10,000$ have been reported for
microcantilevers \cite{Finot}. For a cantilever of length $L =
0.2$mm such a high quality factor would allow the deflection of
the free end due to Einstein - de Haas effect as high as a few
tens of a micrometer.

\section{Acknowledgements}

This work has been supported by the NSF Grant No. DMR-0703639.

\end{document}